\documentclass[twocolumn,pra,aps,10pt,longbibliography]{revtex4-1}

\usepackage{amsmath}
\usepackage{graphicx}

\usepackage{soul,xcolor}
\setstcolor{red}

\usepackage{url}
\usepackage[colorlinks]{hyperref}

\hypersetup{
 plainpages=true,
 breaklinks=true,
 hypertexnames=false,
 pageanchor=true,
 colorlinks=true,
 linkcolor={blue},
 citecolor={red},
 urlcolor={violet},
 anchorcolor={black}
 }

\newcommand{\ket}[1]{|{#1}\rangle}
\newcommand{\bra}[1]{\langle{#1}|}
\newcommand{\braket}[2]{\langle{#1}|{#2}\rangle}

\newcommand{\e}[1]{e^{#1}}

\newcommand{\cc}{_\mathrm{cc}}
\newcommand{\rev}{^\mathrm{(r)}}
\newcommand{\omt}{\omega_\mathrm{target}}
\newcommand{\Omax}{\Omega_{\max}}
\newcommand{\Dedge}{\Delta_\mathrm{edge}}
\newcommand{\Ptrans}{P_\mathrm{trans}}

\DeclareMathOperator{\sinc}{sinc}

\begin{document}
\title{Coherent spectral hole burning and qubit isolation \\ by stimulated Raman adiabatic passage}
\author{Kamanasish Debnath}
\email[e-mail:]{kamanasish.debnath@phys.au.dk}
\affiliation{Department of Physics and Astronomy, Aarhus University, Ny Munkegade 120, DK-8000, Aarhus C, Denmark}
\author{Alexander Holm Kiilerich}
\email[e-mail:]{kiilerich@phys.au.dk}
\affiliation{Department of Physics and Astronomy, Aarhus University, Ny Munkegade 120, DK-8000, Aarhus C, Denmark}
\author{Albert Benseny}
\email[e-mail:]{abc@phys.au.dk}
\affiliation{Department of Physics and Astronomy, Aarhus University, Ny Munkegade 120, DK-8000, Aarhus C, Denmark}
\author{Klaus M{\o}lmer}
\email[e-mail:]{moelmer@phys.au.dk}
\affiliation{Department of Physics and Astronomy, Aarhus University, Ny Munkegade 120, DK-8000, Aarhus C, Denmark}
\date{\today}

\begin{abstract}
We describe how stimulated Raman adiabatic passage (STIRAP) can be applied to create spectral holes in an inhomogeneously broadened system.
Due to the robustness of STIRAP, our proposal guarantees high flexibility and accuracy  and, at variance with traditional spectral hole burning techniques, it may require substantially less time resources since it does not rely upon the spontaneous decay of an intermediate excited state.
We investigate the effects on the scheme of dephasing and dissipation as well as of unintentional driving of undesired transitions due to a finite splitting of the initial and target state.
Finally, we show that the pulses can be reversed to create narrow absorption structures inside a broad spectral hole, which can be used as qubits for precise quantum operations on inhomogeneously broadened few-level systems.
\end{abstract}
\maketitle

\section{Introduction}

Within the last decade, we have seen a surge of interest in utilizing solid state dopants like rare earth ions in crystals~\cite{Casabone2018, PhysRevB.77.125111, Gobron:17, Mitsunaga:91}, nitrogen vacancy (NV) centers in nanodiamonds~\cite{Kaupp2016, Weber8513} and quantum dots in nanoscale semiconductors~\cite{PhysRevLett.119.143601} for quantum operations.
Experimentally, there has been significant progress leading to single photon sources~\cite{PhysRevLett.85.290}, quantum memories~\cite{PhysRevLett.110.250503}, individual addressing of ions~\cite{Kaupp2016}, observation of ultra slow group velocities of light~\cite{PhysRevLett.88.023602} and much more.
However, due to their complex structure and high density, all of these systems suffer from vast intrinsic inhomogeneous broadening~\cite{doi:10.1063/1.4913428, PhysRevLett.117.250504, PhysRevB.70.214116}.
The broadening is predominantly observed in the excited state since it interacts strongly with the surrounding crystal medium which leads to a shift in the transition frequencies.

\begin{figure}
\centering
\includegraphics[width=1.0\columnwidth]{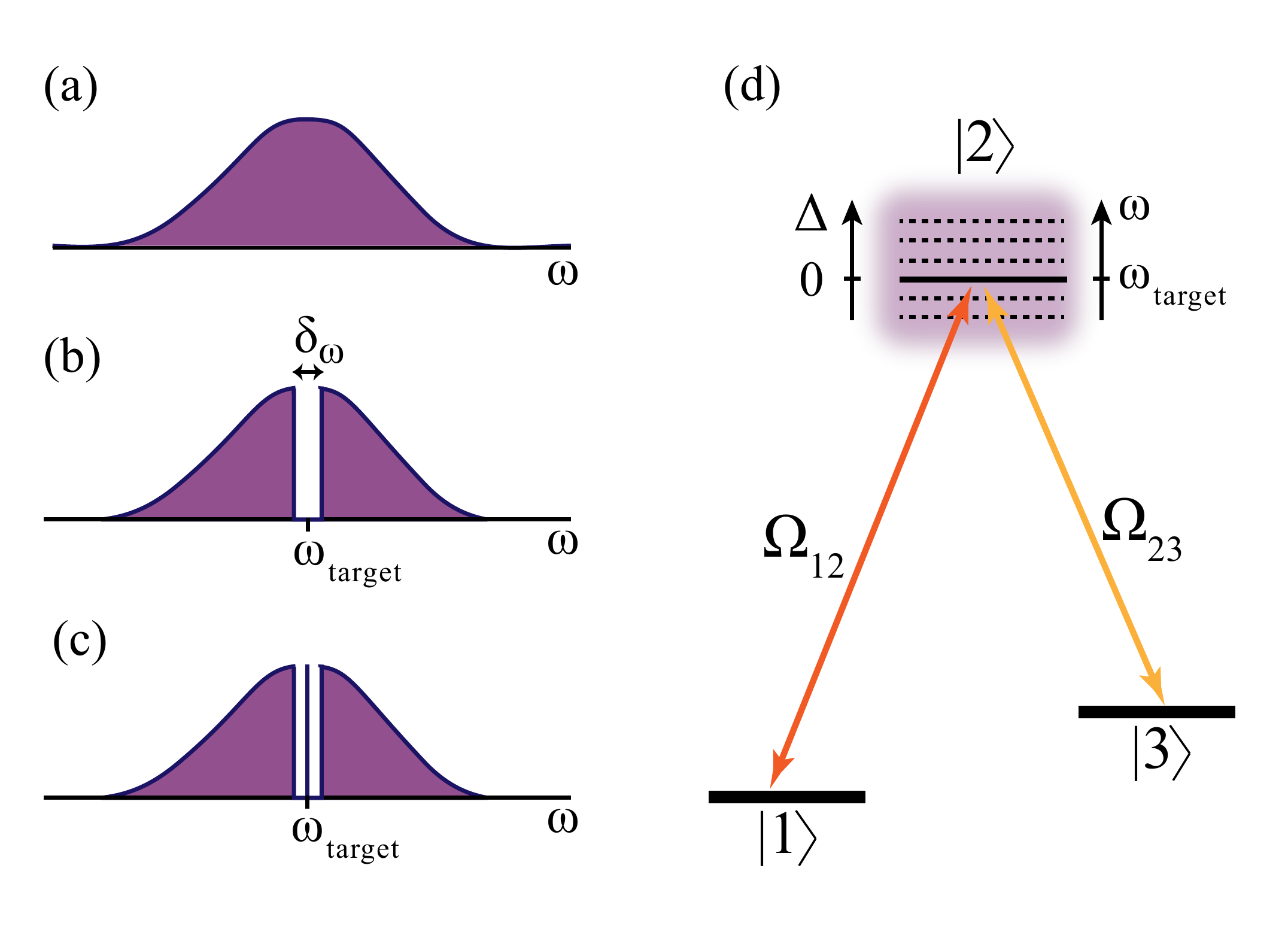}
\caption{Schematic representation of an idealized hole burning process using STIRAP.
(a) Absorption profile of the $\ket{1}\leftrightarrow\ket{2}$ transition.
(b) Spectral hole of width $\delta_\omega$ around the central frequency $\omt$ after application of STIRAP.
(c) Isolated (qubit) structure inside the spectral hole.
(d) Level diagram and couplings commonly used in STIRAP. The dotted lines represent the excited state $\ket{2}$ for different ions and the inhomogeneous broadening is illustrated by the shaded region. The bold line represents the ions for which the pulses are resonant, i.e., for which $E_2 - E_1= \hbar\omt$.
}
\label{fig:sketch}
\end{figure}

One well known method to combat this inhomogeneous broadening is to employ spectral hole burning~\cite{PhysRevB.47.14741, PhysRevB.50.18200}.
This is usually carried out by illuminating monochromatic (laser) light on the sample and sweeping the frequency across a range where the hole is desired.
This excites near resonant ions from the ground state to an excited state.
A subsequent decay to a different ground state, not coupled by the laser, then leads to a hole in the spectrum around the chosen frequencies.
This scheme is time consuming, operating on a time scale defined by the lifetime of the excited state.

In this paper, we propose to utilize stimulated Raman adiabatic passage (STIRAP)~\cite{doi:10.1063/1.458514, Goldner_1994} to create spectral holes and to establish isolated peaks in the inhomogeneously broadened absorption spectrum of the dopant ions in a crystal, as illustrated in Fig.~\ref{fig:sketch}(a-c).
STIRAP constitutes a highly efficient technique to transfer population between internal levels of an ion or a molecule without populating the intermediate levels of the Raman transition.
It is robust to variations in the applied laser fields, flexible in choosing the target level, and, as long as the adiabaticity condition is fulfilled, it yields high fidelities~\cite{RevModPhys.89.015006}.
STIRAP has been realized in a diverse range of settings from trapped ions~\cite{S_rensen_2006,PhysRevLett.115.053003}, to chiral molecules~\cite{PhysRevLett.71.3637} and even for quantum computations~\cite{PhysRevA.73.042321, PhysRevA.88.010303}.
For detailed reviews of this method see~\cite{Bergmann2015, RevModPhys.89.015006}.

We consider an ensemble of $\Lambda$ type systems [see Fig.~\ref{fig:sketch}(d)] with a broad, inhomogeneous distribution [see Fig.~\ref{fig:sketch}(a)] of the transition frequencies to the excited state $\ket{2}$.
In our scheme, spectral hole burning is achieved by adiabatically transferring the population from state $\ket{1}$ to state $\ket{3}$ for those ions with resonance frequencies around a predefined frequency $\omt$.
Despite the robustness of STIRAP against parameter variations, the population transfer fidelity drops if the detuning [$\Delta$ in Fig.~\ref{fig:sketch}(d)] is too large.
We shall show that this defines a sharp cut-off at $\omt\pm \delta_\omega/2$, yielding a spectral hole with well-defined edges where, as illustrated in Fig.~\ref{fig:sketch}(b), every ion has been transferred to $\ket{3}$.
The width $\delta_\omega$ of this spectral hole can be tuned by the pulse parameters and under ideal conditions one can engineer a spectral hole with a completely flat base.

In addition to coherent spectral hole burning, our protocol offers the possibility to \textit{unburn} a part of the hole by adding a second set of reversed pulses with different parameters, thereby preparing a group of emitters which absorb light within a narrow frequency interval inside the wider spectral hole, as illustrated in Fig.~\ref{fig:sketch}(c).
Such a reversed STIRAP (rSTIRAP) protocol can be used to isolate well-defined ions which may serve as qubits within a broad ensemble.
In principle, this can be done at multiple frequency locations to create several qubits which, if geometrically located close to each other, experience dipole-dipole interactions that can facilitate the implementation of controlled NOT (cNOT) gate~\cite{PhysRevA.69.022321}.
Such an architecture will ultimately pave the way towards scalable quantum computing schemes relying on inhomogeneously broadened ensembles of quantum systems~\cite{PhysRevA.75.012304,OHLSSON200271}.

The paper is organized as follows.
We start by presenting a generalized three-level model with a brief description of STIRAP in Sec.~\ref{S2}.
In Sec.~\ref{S3}, we discuss the hole burning protocol and study the parameters that determine the shape and width of the spectral hole.
Furthermore, we evaluate the influence of decoherence channels and off-resonant cross coupling on our scheme.
Then, in Sec.~\ref{S4}, we introduce and characterize a reversed STIRAP process (rSTIRAP) which returns the population to $\ket{1}$ for a selected frequency range.
Finally, we conclude with an outlook in Sec.~\ref{S5}.

\section{Physical system \label{S2}}
We consider a collection of three-level ions in a $\Lambda$ configuration [Fig.~\ref{fig:sketch}(d)] which suffer from intrinsic inhomogeneous broadening as shown by the schematic representation of the absorption spectra on its $\ket{1}\leftrightarrow \ket{2}$ transition in Fig.~\ref{fig:sketch}(a).
This can be any inhomogeneous system, as for example, NV centers in nanodiamonds or a given species of rare earth ion doped in a host crystal like Eu$^{3+}$ ions in Y$_2$O$_3$~\cite{Casabone2018} or Y$_2$SiO$_5$~\cite{Gobron:17} and Nd$^{3+}$ ions in YVO$_4$~\cite{PhysRevB.77.125111}.

\subsection{Three-level model}
The two transitions $\ket{1}\leftrightarrow \ket{2}$ and $\ket{2}\leftrightarrow \ket{3}$ are driven with (time dependent) Rabi frequencies $\Omega_{12}(t)$ and $\Omega_{23}(t)$.
We drive the $\ket{1}\leftrightarrow \ket{2}$ transition with frequency $\omt$ (around which the spectral hole is desired) and assume the two photon resonance condition, which fixes the driving frequency on the $\ket{2}\leftrightarrow \ket{3}$ transition.
Due to the inhomogeneous broadening, each ion perceives a different detuning
$\Delta_i = \omt - \omega_{12}^{(i)}$, where $\omega_{12}^{(i)}$ is the resonance frequency on the $\ket{1}\leftrightarrow \ket{2}$ transition of the $i^{th}$ ion .
For simplicity of notation we will omit the index on $\Delta_i$ below.
Assuming the ions to be non interacting, they each evolve independently according to a Hamiltonian ($\hbar= 1$)
\begin{align}
H &= \frac{\Omega_{12}(t)}{2}\Big(\ket{1}\bra{2} + \ket{2}\bra{1} \Big) + \frac{\Omega_{23}(t)}{2}\Big(\ket{2}\bra{3} + \ket{3}\bra{2} \Big) \nonumber \\ &\qquad + \Delta \ket{2}\bra{2} .
\label{E1}
\end{align}
In realistic settings, the excited state $\ket{2}$ has a finite lifetime and the ensemble suffers from dephasing due to crystal imperfections and stray magnetic fields.
To assess the effects of such mechanisms on our proposal, we describe the evolution of our system by a Lindblad master equation,
\begin{align}
\dot{\rho} &= -i[H,\rho] + \gamma_{21}\mathcal{D} \big[\ket{1}\bra{2}\big]\rho + \gamma_{23}\mathcal{D} \big[\ket{3}\bra{2}\big]\rho
\nonumber \\ & \qquad + \Gamma\mathcal{D}\big[\ket{2}\bra{2}-\ket{1}\bra{1}-\ket{3}\bra{3}\big]\rho,
\label{E2}
\end{align}
where we apply the superoperator $\mathcal{D}[\hat{\mathcal{O}}]\rho = \hat{\mathcal{O}}\rho\hat{\mathcal{O}}^{\dagger} - \frac{1}{2}\{\hat{\mathcal{O}}^{\dagger}\hat{\mathcal{O}}, \rho\}$ to describe spontaneous decay with rates $\gamma_{21}$ and $\gamma_{23}$ on the two optical transitions, and dephasing at a rate $\Gamma$ of the excited state relative to the two ground states.

\begin{figure*}
\centering
\includegraphics[width=\textwidth]{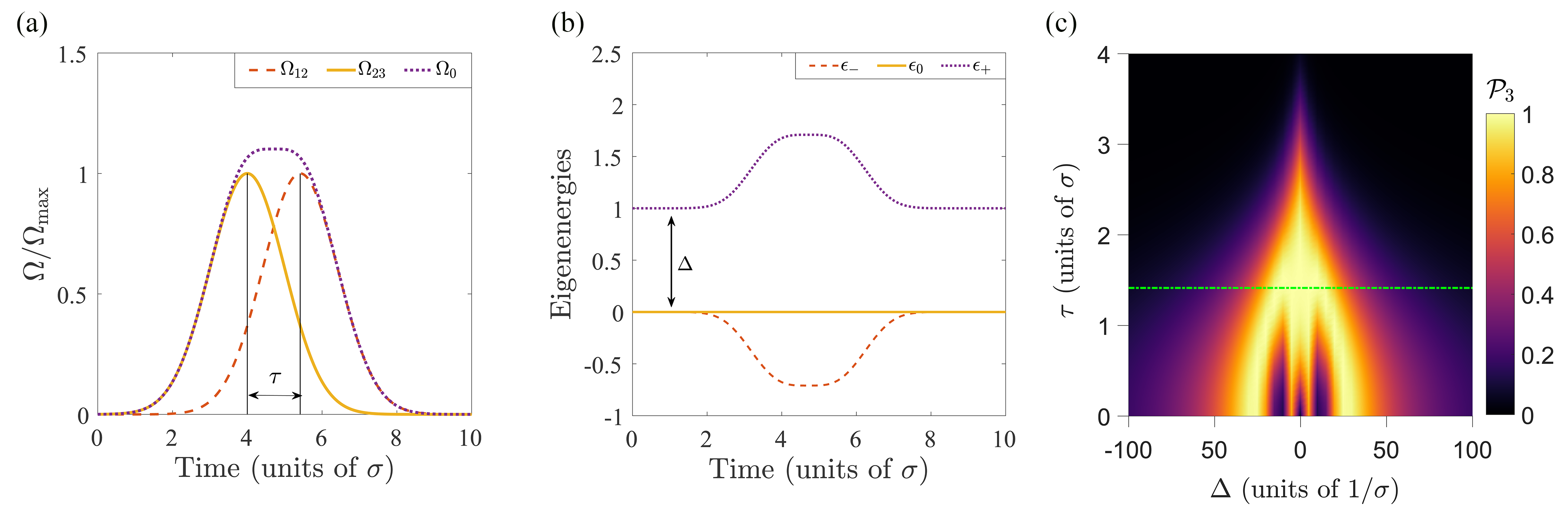}
\caption{STIRAP protocol.
(a) Rabi frequencies during the protocol with the delay $\tau$ indicated.
(b) Spectrum of the Hamiltonian for positive detuning $\Delta$ during the protocol.
(c) Final population $\mathcal{P}_3$ in state $\ket{3}$ as a function of the detuning $\Delta$ and pulse delay $\tau$. The chosen (optimal) time delay $\tau= \sqrt{2}\sigma$, used in the rest of the article is indicated with a green dot-dashed line. Results are shown for $\Omax = 10\sigma^{-1}$ and $\gamma_{21}= \gamma_{23} = \Gamma = 0$.
}
\label{F2}
\end{figure*}

\subsection{STIRAP scheme}

The Hamiltonian~\eqref{E1} has one zero eigenvalue ($\varepsilon_0=0$), with corresponding eigenstate
\begin{equation}
\ket{D} = \cos \theta \ket{1} - \sin\theta\ket{3}.
\label{E3}
\end{equation}
The other eigenvalues are
\begin{equation}
\label{eq:bright_energies}
\varepsilon_\pm = \frac{1}{2} \left(\Delta \pm \sqrt{\Delta^2 + \Omega^2_0} \right)
\end{equation}
and their corresponding eigenstates are
\begin{align}
\label{E4p}
\ket{+} &= \sin\theta \sin\phi\ket{1} + \cos\phi\ket{2} + \cos\theta \sin\phi\ket{3} ,
\\
\label{E4m}
\ket{-} &= \sin\theta \cos\phi\ket{1} - \sin\phi\ket{2} + \cos\theta \cos\phi\ket{3} ,
\end{align}
where
\begin{align}
\tan \theta&= \frac{\Omega_{12}}{\Omega_{23}},
\qquad
\tan 2\phi= \frac{\Omega_0}{\Delta},
\end{align}
with
\begin{equation}
\Omega_0=\sqrt{\Omega^2_{12} + \Omega^2_{23}}.
\end{equation}

Clearly, $\ket{D}$ is a non radiative dark state since it involves only the ground states $\ket{1}$ and $\ket{3}$ with a mixing angle $\theta$, which depends on the two Rabi frequencies.
The idea in STIRAP is to adiabatically follow this dark state while varying $\theta$ between
$\theta= 0$ ($\Omega_{23}\gg\Omega_{12}$, $\ket{D}=\ket{1}$) and
$\theta= \pi/2$ ($\Omega_{23}\ll\Omega_{12}$, $\ket{D}=-\ket{3}$).
This will effectively transfer all the populations from $\ket{1}$ to $\ket{3}$.
The introduction of a time delay $\tau > 0$ between the pulses ensures that the $\Omega_{23}$ pulse arrives before the $\Omega_{12}$ pulse.
This constitutes the core mechanism in STIRAP --- the first pulse opens an energy gap in the spectrum, allowing an adiabatic transfer when the second pulse arrives.

The basic requirement is thus that the two pulses overlap while the mixing angle $\theta$ is varied.
However, as long as the adiabaticity condition is satisfied (see below), the exact shape of the pulses is immaterial.
Hence, for simplicity, we shall consider Gaussian pulses with identical strengths $\Omax$ and widths $\sigma$, as shown in Fig.~\ref{F2}(a),
\begin{align}
\label{eq:Omega12}
\Omega_{12}(t) &= \Omax \exp \left( - \frac{(t-\tau/2)^2}{2 \sigma^2} \right) ,
\\
\label{eq:Omega23}
\Omega_{23}(t) &= \Omax \exp \left( - \frac{(t+\tau/2)^2}{2 \sigma^2} \right) .
\end{align}

\begin{figure*}
\centering
\includegraphics[width=\textwidth]{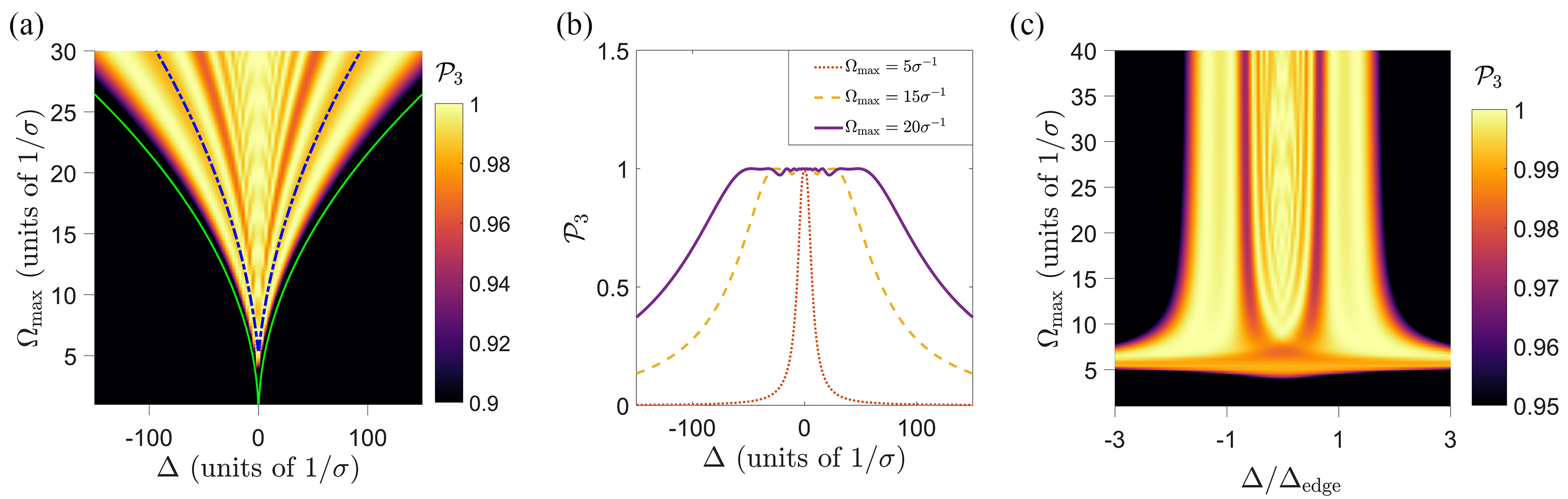}
\caption{ Spectral hole burning.
(a) $\mathcal{P}_3$ as a function of pulse delay and detuning.
The blue, dot-dashed line corresponds to $\pm \Dedge$ defined in Eq.~\eqref{AE1}.
The bold, green line corresponds to $\pm \Omega_0^2 / 16\dot{\theta}$.
(b) Spectral hole profile ($\mathcal{P}_3$) as a function of $\Delta$ for selected values of $\Omax$.
(c) $\mathcal{P}_3$ as a function of $\Omax$ and detuning normalized to $\Dedge$.}
\label{F3}
\end{figure*}

\subsection{Adiabaticity and optimal time delay}

In order to ensure an adiabatic transfer, the local adiabaticity condition must hold throughout the entire process.
Namely,
\begin{equation}
| \varepsilon_0 - \varepsilon_\pm | \gg | \braket{\pm}{\dot{D}} | ,
\label{E5}
\end{equation}
which in the absence of single photon detuning ($\Delta=0$) becomes
$\Omega_0 \gg |\dot{\theta}|$~\cite{RevModPhys.89.015006}.
This condition guarantees that there are no diabatic transitions out of the dark state $\ket{D}$, and can be fulfilled by a sufficiently slow process (having a slow rate of change of $\theta$) or by a large enough gap (larger Rabi frequencies in $\Omega_0$).

In the next section we will see how the presence of detuning due to inhomogeneity affects this condition, as we are interested in engineering a protocol which attains adiabaticity for a tunable range of frequencies $\delta_\omega$ around $\omt$ ($\Delta=0$).
We shall take the amplitude $\Omax$ of the pulses as our main control parameter, and in the following determine an optimal value of the time delay $\tau$ between the pulses.

We simulate~\cite{JOHANSSON20131234} the STIRAP process by considering the ions to be initially in $\ket{1}$ and apply the pulses shown in Fig.~\ref{F2}(a).
The eigenvalues of the Hamiltonian in the presence of single photon detuning $\Delta$ are plotted in Fig.~\ref{F2}(b).
The final population $\mathcal{P}_3$ in $\ket{3}$ is shown in Fig.~\ref{F2}(c) as a function of the single-photon detuning $\Delta$ and pulse delay $\tau$.
For $\tau \ll \sigma$, both pulses are switched on almost simultaneously and both $\ket{1}$ and $\ket{3}$ are coupled to $\ket{2}$.
This leads to Rabi oscillations between the different states leaving population in the excited state at the final time.
However, for certain values of the detuning (around $|\Delta| \simeq 25 \sigma^{-1}$), perfect transfer into $\ket{3}$ is achieved.
These points represent two-photon resonant Raman processes and, unlike the STIRAP process, they are highly dependent on the Rabi frequencies and detuning.
For $\tau \gg \sigma$, the process breaks down due to the complete temporal separation of the two pulses.
The first pulse ($\Omega_{23}$) has no effect since all the population is in $\ket{1}$ while it is on, and the second pulse leads simply to Rabi oscillations between $\ket{1}$ and $\ket{2}$.
In-between these extreme limits, however, there is a regime where STIRAP works perfectly.
We fix $\tau=\sqrt{2}\sigma$, indicated by the green, dot-dashed line in panel Fig.~\ref{F2}(c), which has been found to be an optimal value for the delay in similar studies~\cite{RevModPhys.70.1003}.

\section{Hole burning \label{S3}}

We will now describe how STIRAP can be used for hole burning in an inhomogeneously broadened sample.
We initially neglect the effects of dephasing and decay ($\gamma_{21} = \gamma_{23} = \Gamma=0$ in the master equation~\eqref{E2}) but we shall discuss their implications for our results in Sec.~\ref{sec:decoherence}.

\subsection{Spectral hole profile}
Figure~\ref{F3}(a) shows $\mathcal{P}_3$ after the STIRAP pulses as a function of $\Delta$ and $\Omax$.
We notice that for ions for which the pulses are very off-resonant, STIRAP fails and only ions with $\Delta\simeq 0$ are transferred to state $\ket{3}$.
However, as the strength of the pulses increases, the protocol becomes robust in a wider range of detunings, 
and the width $\delta_\omega$ of the spectral hole increases.
This is due to a growing energy gap between the dark state and the relevant bright state (see Eqs.~\eqref{eq:bright_energies}, and our discussion below).

The spectral profile of the hole is shown in Fig.~\ref{F3}(b) for different values of $\Omax$.
From these plots, it is evident that while small oscillations appear in the base of the spectral hole, with higher values of $\Omax$ one can achieve a spectral hole with a fairly flat base and sharply defined edges beyond which the population in $\ket{3}$ decreases to zero.
It is noteworthy that the robustness of the STIRAP guarantees that the excited state $\ket{2}$ is never populated during this population transfer, which was also observed in our simulations.

We can address the width of the spectral hole by considering the effects of the single-photon detuning $\Delta$ on the adiabaticity condition.
In the presence of finite $\Delta$, one of the bright states ($\ket{+}$ for $\Delta>0$ and $\ket{-}$ for $\Delta<0$) decouples from the dark state $\ket{D}$ with an energy gap larger than $\Delta$.
Therefore, it is sufficient to consider only the adiabaticity condition involving the other eigenstate.
We focus our discussion on the case where $\Delta > 0$ and study the coupling with $\ket{-}$, and note that the reverse situation ($\Delta < 0$, coupling to $\ket{+}$) produces the expected symmetrical results.
The STIRAP adiabaticity condition~\eqref{E5} yields approximately $|\Delta| \ll \Omega_0^2 / 8\dot{\theta}$ which already gives the scaling of the width with the pulse parameters.
As seen from the green line in Fig.~\ref{F3}(a), defining the edge of the hole at $ \pm \Omega_0^2 / 16\dot{\theta}$ fits the numerical data with good agreement.

We may apply a simple model which captures the essential features of the STIRAP transition, and allows analytic insight regarding the width and detailed profile of the spectral hole and its dependence on the pulse parameters.
The model assumes that $\Omega_{0}(t)$ and $\phi(t)$ are constant while the mixing angle $\theta(t) = {\pi t}/{2 T}$ varies linearly between 0 and $\pi/2$ during a time $T$.
This approximates well the Gaussian pulses in Eqs.~(\ref{eq:Omega12}-\ref{eq:Omega23}) in the time interval between their maxima where the main part of the STIRAP process happens.

\begin{figure}
\centering
\includegraphics[width=0.38\textwidth]{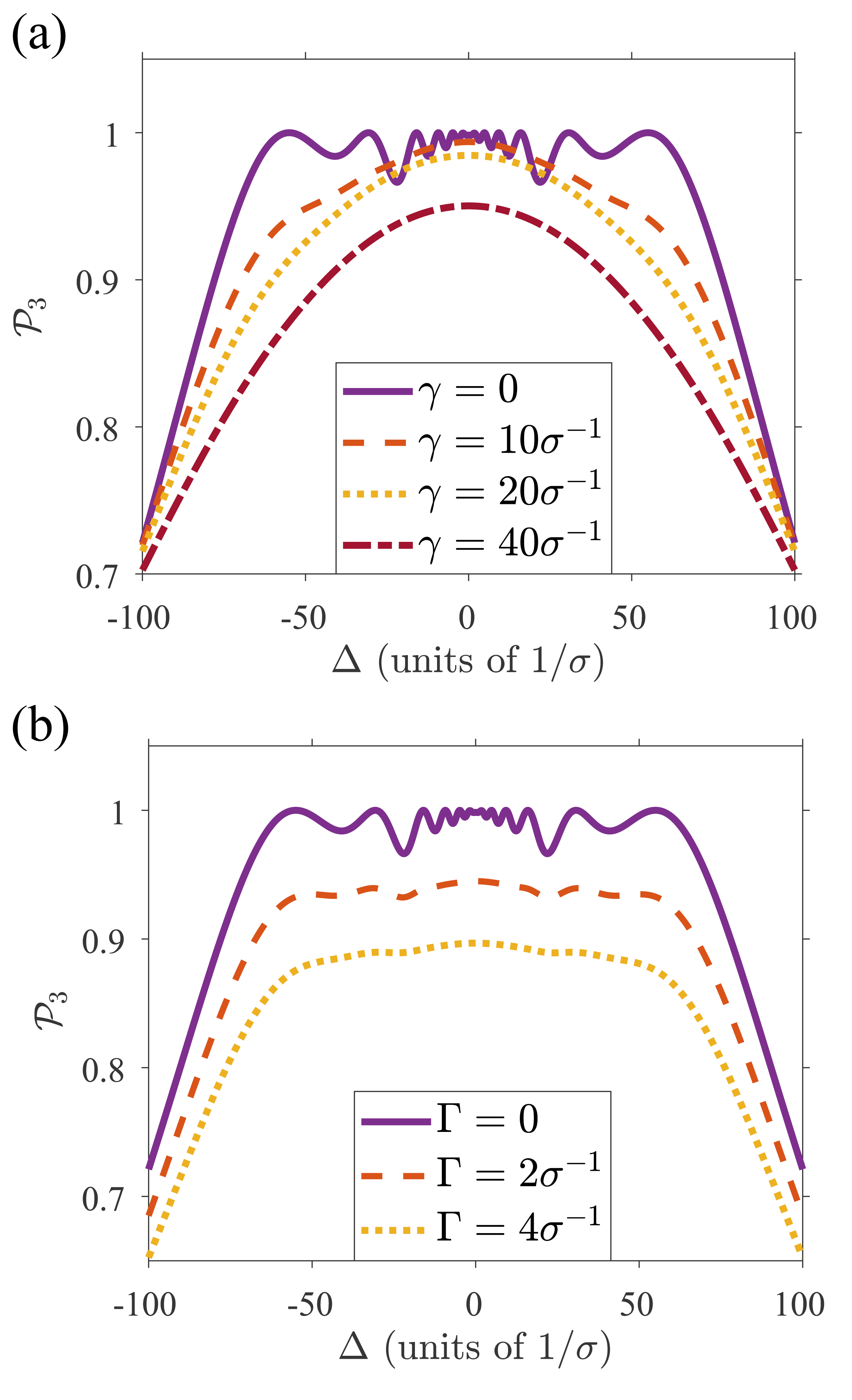}
\caption{Effects of decay and dephasing.
Spectral hole profile ($\mathcal{P}_3$) as a function of $\Delta$ for different values of (a) the excited state decay rates $\gamma_{21}=\gamma_{23}=\gamma$ and (b) the excited state dephasing rate $\Gamma$.
Results are shown for $\Omax = 20\sigma^{-1}$.}
\label{fig:decoherence}
\end{figure}

In absence of dephasing and decay, a perturbative treatment yields the probability for a non-adiabatic transition out of the dark state~\cite{Messiah_1961},
\begin{align}
\Ptrans &\simeq \frac{ \left| \int_{0}^{T} \braket{-}{\dot D} \exp\left[-{i \int_{0}^{t} \delta_\epsilon \, dt^\prime }\right] dt \right|^2}{ \left| \int_{0}^{T} \braket{-}{\dot D} dt \right|^2}.
\end{align}
Under the above conditions, both the energy gap $\delta_\epsilon \equiv (\varepsilon_{0} - \varepsilon_{-})$ and the non-adiabatic coupling $\braket{-}{\dot D}$ remain constant,
and we obtain an analytical estimate for the final population of state $\ket{3}$
\begin{align}
\label{eq:sinc}
\mathcal{P}_3 = 1-\Ptrans
= 1-\sinc^2 \left(\frac{\pi \delta_\epsilon}{4 \dot\theta}\right) ,
\end{align}
where $\sinc(x) = \sin(x)/x$.
Note that
$\delta_\epsilon = ( \sqrt{\Delta^2 + \Omega^2_0} - \Delta ) / 2$
decreases monotonically as $\Delta$ increases.
The edge of the spectral hole is defined by the first zero of the $\sinc$ function, which occurs when
$\pi\delta_\epsilon/4\dot\theta = \pi$,
yielding
\begin{gather}
\Dedge = \frac{\Omega_{0}^2}{16 \dot\theta} - 4 \dot\theta ,
\label{AE1}
\end{gather}
and a corresponding hole width of $\delta_\omega = 2\Dedge$.
For $|\Delta| < \Dedge$, $\mathcal{P}_3$ is close to 1, giving a flat plateau for the spectral hole ($\sinc^2(x) < 0.05$ for $x>\pi$).
For $|\Delta| > \Dedge$, $\mathcal{P}_3$ rapidly decreases and for large $\Delta$, Eq.~\eqref{eq:sinc} can be approximated as $\mathcal{P}_3\simeq(\pi^2\Omega^4_0/768\dot{\theta}^2\Delta^2) + \mathcal{O}[1/\Delta^4]$.

In order to apply this expression to the Gaussian pulses, we approximate the parameter values by those at the instant between the two pulses (i.e., Eqs.~(\ref{eq:Omega12}-\ref{eq:Omega23}) at $t=0$), giving
$\Omega_{0} \simeq \sqrt{2} e^{-\tau^2/8 \sigma^2} \Omax$ and $\dot\theta \simeq \tau/2 \sigma^2$.
With these approximations, $\Dedge$ is shown with a blue dot-dashed line in Fig.~\ref{F3}(a) which fits the numerical data with reasonable agreement.

We can also use Eq.~\eqref{eq:sinc} to understand the structure of the base of the hole as seen in Fig.~\ref{F3}(b).
The oscillatory structure is defined by the tail of the $\sinc$ function beyond its first zero.
The maximum value of the argument is $\pi\Omega_{0}/8\dot\theta$ (at $\Delta=0$) which dictates the number of oscillations and how flat the base appears (how far out the $\sinc$ tail it starts).
Figure~\ref{F3}(c) shows $\mathcal{P}_3$ as a function of $\Delta/\Dedge$ and pulse intensity $\Omax$.
This allows to see the generality of this simple model in predicting the cut off, and to visualize the structures of the spectral hole with its increasing flatness as $\Omax$ increases.

\begin{figure}
\centerline{\includegraphics[width=0.36\textwidth]{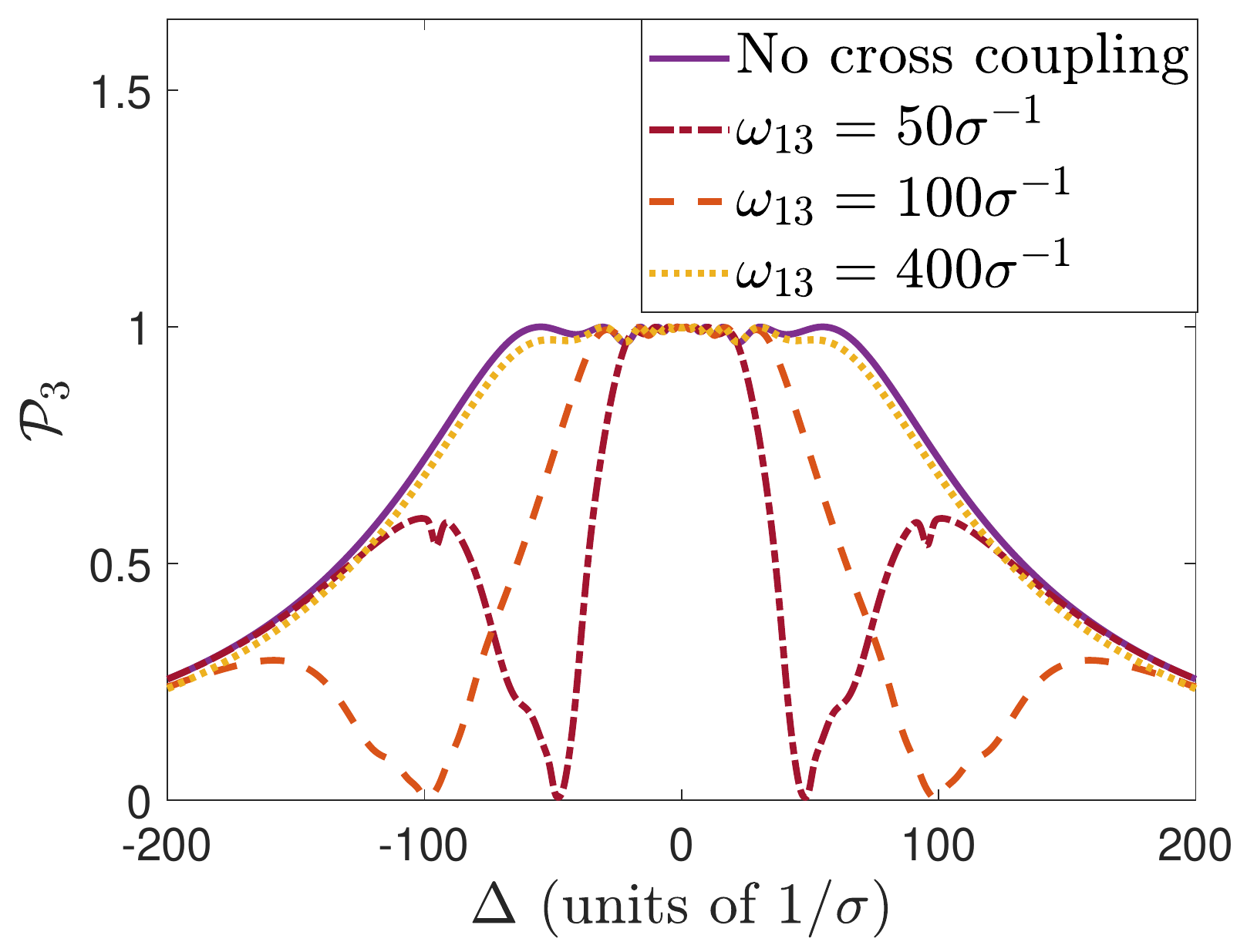}}
\caption{
Spectral hole profile ($\mathcal{P}_3$) for different values of the splitting $\omega_{13}$ between the stable state $\ket{1}$ and $\ket{3}$. The solid curve assumes that each laser couples only to the desired transition in the ions.
Results are shown for $\Omax = 20\sigma^{-1}$ and $\Gamma= \gamma_{21}= \gamma_{23}= 0$.}
\label{fig:crossCoupling}
\end{figure}

\subsection{Effects of decay and dephasing}
\label{sec:decoherence}

While the results presented above signify that under ideal conditions STIRAP allows well defined and broad holes to be burned in an inhomogeneous profile, any realistic setup will experience experimental or fundamental limitations.
In the following, we study the effects of decay and dephasing on the hole burning process.

We show in Fig.~{\ref{fig:decoherence}}(a) the profile of the spectral hole for different decay rates of the excited state ($\gamma_{21}=\gamma_{23}\equiv \gamma\geq 0$).
We find that as the rate of decay increases, the flat profile of the spectral hole is lost.
Moreover, for very large values of $\gamma$ even the high fidelity at resonance is lost.

In Fig.~{\ref{fig:decoherence}}(b), it is seen that while our protocol is robust in the absence of non-radiative dephasing ($\Gamma\geq 0$) of the excited state, as the dephasing increases, the final population $\mathcal{P}_3$ in state $\ket{3}$ starts decreasing.
At variance with the effects of decay, however, the flatness of the hole is preserved as the fidelity decreases globally.

Our results shows thus, that for systems with large environmental couplings, the STIRAP hole burning protocol becomes inefficient.
However, realistic dephasing and decay rates are typically much smaller than the values considered in Fig.~{\ref{fig:decoherence}}~\cite{Casabone2018}, ensuring good performance for ions in quantum information processing tasks.

\subsection{Effects of off-resonant cross coupling}

\begin{figure*}
\centering
\includegraphics[width=\textwidth]{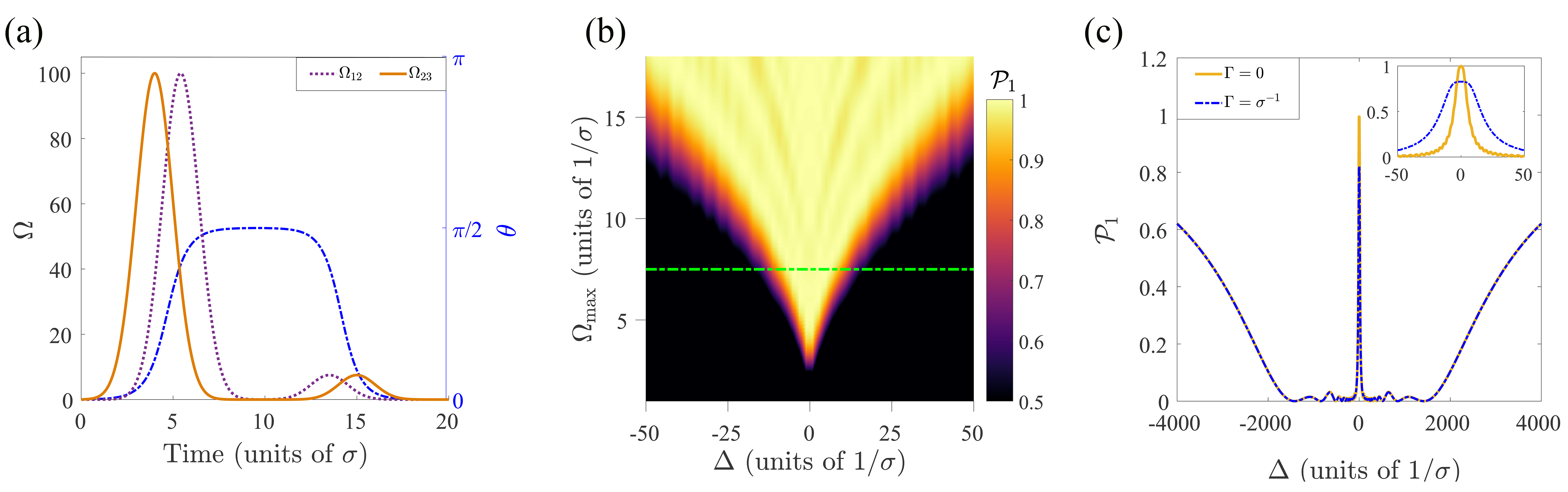}
\caption{
Qubit isolation around a single target frequency $\omt^{(n)}$ coinciding with the centre of the spectral hole.
(a) Pulse Rabi frequencies $\Omega_{12}(t)$ and $\Omega_{23}(t)$ and mixing angle $\theta(t)$ during the combined STIRAP and by rSTIRAP protocols.
(b) Final population $\mathcal{P}_1$ in state $\ket{1}$ as a function of the detuning and pulse amplitude of rSTIRAP.
(c) Final population $\mathcal{P}_1$ in state $\ket{1}$ in the presence and absence of dephasing for the pulses shown in (a). The inset shows a zoom-in on the central structure.
Results are shown for $\gamma_{21} = \gamma_{23} = 0$.
}
\label{fig:qubit}
\end{figure*}
This far we have assumed that each laser couples only to the desired transition in the ions.
For some ions, however, the detuning may be comparable to the splitting between the stable states ($|\Delta|\simeq \omega_{13}$), and one of the STIRAP laser pulses may unintentionally excite the wrong transitions, as described by the contributions to the Hamiltonian,
\begin{align}
\begin{split}
H\cc =
\frac{\Omega_{23}(t)}{2}\Big(\ket{1}\bra{2}\e{-i\omega_{13} t} + \ket{2}\bra{1}\e{i\omega_{13} t} \Big)
\\
+ \frac{\Omega_{12}(t)}{2}\Big(\ket{3}\bra{2}\e{i\omega_{13} t} + \ket{2}\bra{3}\e{-i\omega_{13} t} \Big),
\end{split}
\end{align}
that were disregarded in Eq.~\eqref{E1}.

In Fig.~\ref{fig:crossCoupling} it is seen that as long as the level splitting is larger than the desired spectral hole (cf. the curve for $\omega_{13}=400\sigma^{-1}$), the effects of this cross-coupling are insignificant in the intended frequency range.
For smaller level splittings, however, the hole width is reduced, as STIRAP becomes completely ineffective and yields no final population in $\ket{3}$ for the resonances $\Delta = \pm \omega_{13}$ (cf. the dips for the $\omega_{13}=50\sigma^{-1}$ and $100\sigma^{-1}$ curves).
It is thus evident that the energy splitting imposes an upper bound $\delta_{\omega}\lesssim 2\omega_{13}$ to the width of the spectral hole.
Note that a similar restriction applies for conventional hole burning.

\section{Qubit isolation \label{S4}}

We have shown how STIRAP can be utilized to create spectral holes with high fidelity.
In this section, we propose to follow that process with a set of \textit{reversed} STIRAP (rSTIRAP) pulses of lower intensity in order to bring some atoms back into state $\ket{1}$ and create narrow-band peaked structures at target frequencies $\omt^{(n)}$ inside the burned hole.
This can be used, for instance, in quantum information processing to effectively create one or multiple isolated qubits in an inhomogeneous sample~\cite{PhysRevA.75.012304, OHLSSON200271}.

\subsection{Reversed STIRAP}
Assuming ideal conditions, after the initial STIRAP pulses, all ions inside the spectral hole are in state $\ket{3}$.
Therefore, a new set of reversed pulses
 ($\Omega_{12}$ acting before $\Omega_{23}$, i.e., $\tau < 0$ in Eqs.~(\ref{eq:Omega12}-\ref{eq:Omega23}))
will adiabatically bring back a selected band of them to the state $\ket{1}$.
As we discussed in the previous section, the width of the band is tunable through the pulse intensity, and less intense pulses thus allow us to bring back a much narrower band of the spectrum.
The set of STIRAP and rSTIRAP pulses in this process are depicted in Fig.~\ref{fig:qubit}(a), which also shows how the mixing angle $\theta(t)$ (blue, dotted line) varies from $0$ to $\pi/2$ for the STIRAP pulses, followed by its return to $0$ during the rSTIRAP pulses.

Figure~\ref{fig:qubit}(b) shows how the strength of the second set of pulses, $\Omax\rev$, influences the final population in $\ket{1}$ for different values of $\Delta$ (after the hole has been created).
It is seen that, in a similar manner as for the original hole burning, the strength of the rSTIRAP pulses can indeed be used to control the width of the \textit{unburnt} frequency band.
The figure shows also the limit on how narrow these peaks can be created, as dictated by the adiabaticity condition: if $\Omax\rev$ is too small, rSTIRAP fails for \textit{all} values of $\Delta$, resulting in a peak with less than unit population in $\ket{1}$.

\subsection{Isolation profiles}

We show in Fig~\ref{fig:qubit}(c) the profile of a narrow isolated peak inside a broad spectral hole, signifying that the combination of STIRAP and rSTIRAP processes can be successfully employed to create well-defined systems within a broad inhomogeneous ensemble.
The wiggles in the spectral hole, see Fig.~\ref{F3}, can cause a small loss of fidelity of the imprinted structures around the desired qubit.
However, for multi-qubit quantum computing protocols~\cite{OHLSSON200271}, the isolated emitters will dominate over the background.

\begin{figure}
\centering
\includegraphics[width=0.38\textwidth]{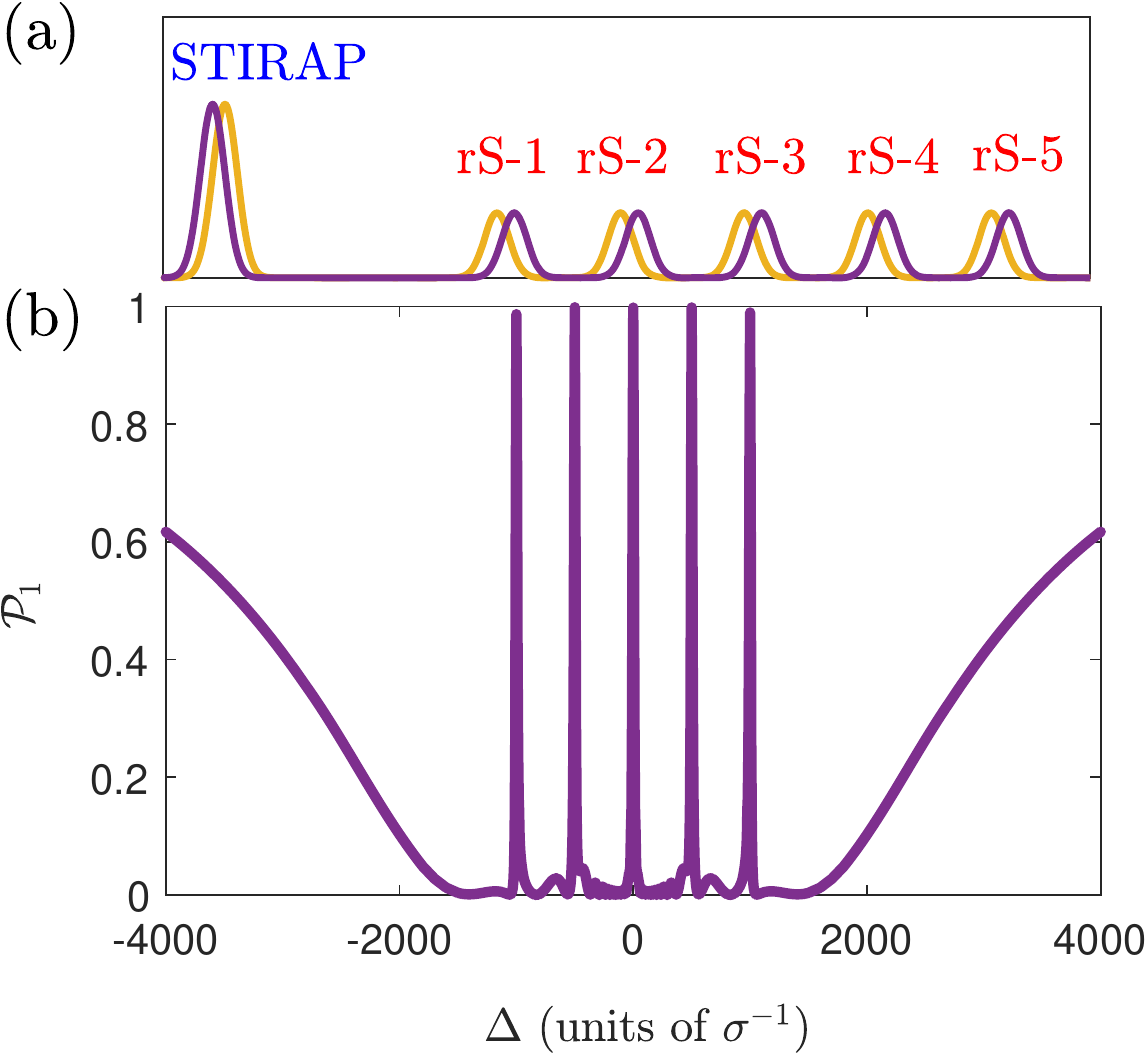}
\caption{
Multiple isolated qubits.
(a) Sketch of the STIRAP pulses and subsequent sets of rSTIRAP pulses with five different target frequencies $\omt^{(n)}=\omt+n\times500\sigma^{-1}$ with $n = -2,-1,0,1,2$.
(b) Final population $\mathcal{P}_1$ in state $\ket{1}$ after the train of pulses.
Results are shown for $\Omax = 100 \sigma^{-1}$, $\Omax\rev = 5 \sigma^{-1}$, $\gamma_{21}= \gamma_{23} = \Gamma = 0$.
}
\label{F6}
\end{figure}

It is possible to isolate multiple qubits inside the same broad spectral hole.
We illustrate this in Fig.~\ref{F6}, where the sequential application of five sets of rSTIRAP pulses with different target frequencies $\omt^{(n)}$ yield five well-defined qubits inside the hole.
While here we have assumed the serial application of the different rSTIRAP pulses, it should also be possible to create an atomic frequency comb~\cite{PhysRevA.79.052329,PhysRevLett.104.040503}, i.e., a number of equidistant isolated qubits, by simultaneously applying the rSTIRAP pulses with an optical frequency comb, see also Ref~\cite{PhysRevA.98.043834} for a related proposal in a Doppler-broadened system.

\section{Outlook \label{S5}}

In this paper, we have shown how STIRAP can be used for coherent spectral hole burning with high fidelity and better control of the shape of the spectral hole.
We derived an analytical expression that estimates the width of the hole as a function of the pulse intensity and emphasize how a hole with well-defined edges may be obtained due the adiabaticity requirement in the STIRAP scheme.
Moreover, we proposed to apply additional sets of weaker STIRAP pulses to isolate qubits inside the spectral hole.
STIRAP is highly flexible in targeting any state and the entire process requires just two Gaussian pulses.

Our proposal is robust against variations in pulse parameters as long as the adiabaticity condition is fulfilled and against moderate decoherence in the excited state.
For clarity of presentation, we focused our attention on Gaussian pulses.
However, in a future study it would be interesting to consider the possibilities offered by more complex pulse shapes which may deliver spectral holes with sharper edges and other desirable features.

The parameters considered in this paper are compatible with the state-of-the-art developments in the experimental domain.
For instance, for Eu$^{3+}$ ions doped in an Y$_2$O$_3$ crystal~\cite{Casabone2018}, the two stable states in our model may be the hyperfine states $^7F_0$ and $^7F_2$ with transition frequencies $\omega_{12}= 5.167 \times 10^{14}$ Hz and $\omega_{23}= 4.903 \times 10^{14}$ Hz to the excited state $^5D_0$.
The excited state is known to have an inhomogeneous broadening of $22$ GHz ~\cite{Casabone2018}.
Assuming that one desires to create a spectral hole of width $\delta_\omega \sim 1$ MHz with isolated (qubit) structures of a few kHz wide inside, the dephasing and decay rates of $\Gamma = 0.785$ kHz and $\gamma = 45$ kHz are small and will not severely affect the efficiency of STIRAP.
In addition, the ground state separation is on the order of $\omega_{13} \simeq 0.3\times 10^{14}$ Hz $\gg \delta_\omega$, implying that off-resonant cross coupling is negligible for these systems.

\section{Acknowledgements}
We would like to thank Signe Seidelin and Bess Fang for discussions on possible candidate systems for our proposal.
K.~D. and K.~M. are supported by the European Union's Horizon 2020 research and innovation program (No. 712721, NanoQtech).
A.~H.~K. and K.~M. acknowledge support from the European Union FETFLAG program, Grant No. 820391 (SQUARE). 
A.~B. and K.~M. are supported by the Villum Foundation.

\bibstyle{apsrev4-1}
\bibliography{cshbaqibs}

\end{document}